\documentstyle[twoside, epsf]{article}

\input ibvs2.sty

\begin{document}
\IBVShead{6xxx}{01 February 2013}

\IBVStitletl{RR Lyrae Stars In The GCVS Observed By The}{Qatar Exoplanet Survey}


\begin{center}
\IBVSauth{Bramich, D.M.$^1$; Alsubai, K.A.$^1$; Arellano Ferro, A.$^2$; Parley, N.R.$^{1,3}$; Collier Cameron, A.$^3$; Horne, K.$^3$; Pollacco, D.$^4$; West, R.G.$^4$}
\end{center}

\IBVSinst{Qatar Environment and Energy Research Institute, Qatar Foundation, Tornado Tower, Floor 19, P.O. Box 5825, Doha, Qatar}
\IBVSinst{Instituto de Astronom\'ia, Universidad Nacional Aut\'onoma de M\'exico, Ciudad Universitaria CP 04510, M\'exico}
\IBVSinst{SUPA, School of Physics and Astronomy, University of St Andrews, North Haugh, St Andrews, Fife, KY16 9SS, UK}
\IBVSinst{Department of Physics, University of Warwick, Coventry, CV4 7AL, UK}

\begintext

\section{Abstract}

We used the light curve archive of the Qatar Exoplanet Survey (QES) to investigate the RR Lyrae variable stars listed in the General
Catalogue of Variable Stars (GCVS). Of 588 variables studied, we reclassify 14 as eclipsing binaries, one as an RS Canum Venaticorum-type
variable, one as an irregular variable, four as classical Cepheids, and one as a type II Cepheid, while also improving their periods.
We also report new RR Lyrae sub-type classifications for 65 variables and improve on the GCVS period estimates for 135 RR Lyrae variables.
There are seven double-mode RR Lyrae stars in the sample for which we measured their fundamental and first overtone periods.
Finally, we detect the Blazhko effect in 38 of the RR Lyrae stars for the first time and we successfully measured the Blazhko period
for 26 of them.

\section{Introduction}

The Qatar Exoplanet Survey (QES; Alsubai et al. 2013) is discovering hot Jupiters (Qatar-1b, Alsubai et al. 2011; Qatar-2b, Bryan et al. 2012)
and aims to discover hot Saturns and Neptunes that transit in front of relatively bright host stars (8-15 mag). The survey operates a
robotic wide-angle multiple-camera system installed at the ``New Mexico Skies'' observing station in southern New Mexico, USA, and it
has been in operation since mid-November 2009. The cameras, which operate without filters for maximum signal-to-noise (S/N),
photometrically survey a target field of $\sim$400 square degrees repeatedly with a cadence of $\sim$10 minutes. Each target field is
followed for $\sim$3-4 months continuously while it is visible at greater than 30\deg \, above the horizon. Each year a new set of target fields are designated. 

The time-series images of each field are processed by a customised data pipeline (Sec.~4 of Alsubai et al. 2013) to calibrate the images, detect objects,
perform astrometry, and extract photometry. Only objects successfully matched with stars in the US Naval Observatory CCD Astrograph Catalog (UCAC3; Zacharias et al. 2010)
are considered further in order to avoid faint stars with very low S/N. A reference image, chosen as a best-seeing high-S/N image from the time series, is subtracted from
each image in the time series using the image subtraction technique to create difference images (Alard \& Lupton 1998; Bramich 2008; Bramich et al. 2013). Photometry is
performed on the difference images using point spread function (PSF) fitting at the object positions with a spatially-variable PSF model. The output of this difference image
analysis (DIA) is a set of object light curves in differential flux units (ADU/s). These light curves are converted to instrumental magnitudes using reference fluxes for each object
as measured on the reference image. The photometric zero point for the reference image is determined using the UCAC3 magnitudes and this is used to calibrate the
light curve magnitudes on an absolute scale with a scatter of $\sim$0.1~mag. The QES light curves are then stored in a data archive system and trend filtering algorithms are applied to them.
However, since the application of trend filtering algorithms to variable star light curves risks distorting their shape, we opted to use the raw
QES light curves from the archive (i.e. before detrending is applied) for the study of the variable stars in this paper.

With a typical photometric precision of $\sim$1-2\% over the magnitude range 8-14, a high temporal cadence ($\sim$10~min) sustained over $\sim$2-7 hours in each 24-hour period, and a
time baseline of $\sim$3-4 months, the QES light-curve archive is a potential gold mine for variability studies. As part of realising the full scientific potential of QES, we have started
to investigate the variable star content of the archive. This short paper is the first in a series reporting our results. Here we investigate known RR Lyrae variables.

\section{Sample Selection}

We cross-matched UCAC3 with the 47969 variable stars in the General Catalogue of Variable Stars (GCVS4 - version 30/04/2013; Samus et al. 2009)
using the CDS X-Match service\footnote{http://cdsxmatch.u-strasbg.fr/xmatch\#tab=xmatch\&}. The cross-match algorithm simply selects any GCVS star entries
within a 5\arcs \, radius of any UCAC3 star. This resulted in 43009 matched entries, of which 42973 are unique. Retaining only
the unique matches and filtering for variable star type, we obtained 6921 UCAC3 stars classified as RR Lyrae variables.

We then searched in the QES light-curve archive for these UCAC3 RR Lyrae stars and found that we had observed 752
objects in this list. We note that any object observed across multiple target fields and/or cameras will have multiple light curves
in the QES archive. Since our analysis requires a reasonable number of data points in each light curve, we rejected light curves with
fewer than 100 data points. Furthermore, due to the faint limit of the QES lying at $\sim$17~mag, we rejected any objects
with UCAC3 aperture magnitudes fainter than 16.5. We were left with 724 objects with 2220 light curves.

We inspected plots of the phased (using the GCVS periods where available) and unphased light curves of our object sample. Since
RR Lyrae variations have typical amplitudes of 0.1-1.3~mag, we could immediately identify 65 objects with multiple light curves where
a subset of the light curves were not showing any variability. This occurs when the QES pipeline misidentifies an object and measures the
wrong star, which tends to happen for relatively crowded objects towards the edge of a detector where camera distortions are not
sufficiently well-modelled in the astrometric solution. For these cases we simply rejected the 143 light curves that failed to show
the variations clearly visible in the remaining light curves for the same object. We also identified 136 objects for which none of
their light curves showed variations above the noise level. We found that this was due either to the objects being very faint and
therefore exhibiting a large scatter in their light curves, or to the object misidentification problem mentioned already. We rejected
these objects from our sample, which left us with 588 photometrically variable objects with 1783 light curves.

\section{Analysis And Results}

Some variables in our data sample do not have GCVS period estimates and/or their GCVS classification as RR Lyrae variables
is uncertain or does not distinguish between fundamental mode and first overtone pulsators. Hence our first step was to 
estimate the variable star periods using our light curve data. We applied the string-length method (Burke, Rolland \& Boy 1970; Dworetsky 1983)
to each of the 1783 light curves in our sample to search for periods in the range 0.1-500~d. For variables with multiple light curves,
we adopted the period derived from the light curve with the best combination of the longest time span, the smallest noise, and the most data points
(all light curve plots in this paper display this ``best'' light curve for clarity). We then phase-folded the light curves with our derived periods,
and we checked the RR Lyrae classification of our variables.

Apart from being able to improve the GCVS periods and classifications for a large number of variable stars, we also found that some
variables in our data sample are not RR Lyrae stars. Consequently we have reclassified these stars using the GCVS
classification system described in Samus et al. (2009)\footnote{See also http://www.sai.msu.su/gcvs/gcvs/iii/vartype.txt}. To aid in our reclassification
efforts, we searched the literature for previous studies of some of these variables. However, a full literature search
for all of the variables in our data sample is outside of the scope of this paper, the purpose of which is to provide a set of
concrete updates to the latest version of the GCVS. Therefore we cannot claim that all of our results are guaranteed to
be new although we are sure that the majority of the information presented in this paper has not previously been reported in the literature.

Before reporting our results, we mention that due to the coarse pixel scale of the QES camera system (9.26 and 4.64~arcsec/pixel for the 200 and 400~mm lenses, respectively),
a relatively high proportion of the variable stars in our sample are likely to be blended with another star. Hence the reference flux for such blended variables as measured
on the reference image is systematically over-estimated which leads to artificially small amplitudes of variation in the corresponding light
curves. Therefore, the amplitudes of our variable star light curves may be systematically too small in a number of cases when compared
to light curves derived from higher resolution imaging data.

\subsection{Stars that are not RR Lyrae variables}

In Table~1, we report the reclassification of 21 variable stars. Our period estimates improve on the GCVS periods in all cases.
We reclassify 14 of these variables as eclipsing binaries, where 13 of these are of the W Ursae Majoris-type. We plot the phased
light curves of the eclipsing binaries in Figure~1 using the best light curve for each variable. Four eclipsing
binaries (3UC191-025421, 3UC192-006598, 3UC247-041882 and 3UC308-105518) clearly show the O'Connell effect in our data, which is characterised by two maxima of
different brightnesses (O'Connell 1951).

We find that the variable star 3UC205-101683, which is listed in the GCVS as an RR Lyrae star of unknown sub-type, is a
known double-lined spectroscopic binary (Mathieu et al. 2003) showing X-ray emission (Belloni, Verbunt \& Mathieu 1998) and classed as an
RS Canum Venaticorum-type variable (van den Berg et al. 2002). We have updated the record for this star in Table 1, quoting the period derived
from our best light curve spanning $\sim$154 days, which is more precise than the photometric periods quoted in the literature. The phased light
curve for this star is shown in Figure~2. We note that the slight variations in the light curve shape and amplitude reported by
van den Berg et al. (2002) are also detected in our light curve.

The variable star 3UC171-023140 is a Herbig Ae/Be star of spectral type B9e (Vieira et al. 2003) that exhibits irregular light variations  
(see Figure~3; Bernhard 2010). We were unable to find periodicity in our light curves for this object. Hence we reclassify this star as an 
irregular variable of early spectral type in Table~1.

We found that four of the variable stars in our sample are most likely classical Cepheids as opposed to RR Lyrae stars. Our new classifications for 
two of these stars (3UC208-318430 and 3UC225-131002) are based only on the period and the shape of the phased light curve (see Figure~4), which
does not definitively distinguish them from other variable types in their period range. Hence our classifications for these two stars are tentative
and marked with a colon ``:'' in Table~1. The variable star 3UC237-053427 was originally classified as a classical Cepheid by Schmidt \& Gross (1990)
and we confirm that it is most likely pulsating in the first overtone mode as indicated by its smaller pulsation amplitude and relatively symmetric
phased light curve. The variable star 3UC268-064903 has also already been classified as a classical Cepheid by Wils, Lloyd \& Bernhard (2006).

We also noticed that the variable star 3UC237-121450 has an unusually long period for an RR Lyrae. A literature search revealed that Wallerstein, Kovtyukh \& Andrievsky (2009) found
this star to be carbon-rich and of relatively high metallicity. These facts lead Andrievsky et al. (2010) to suggest that 3UC237-121450 is more likely to be a short-period
type II Cepheid (or BL Her type variable). We adopt this tentative classification for this star in Table 1 and display the phased light curve in Figure 5.

\begin{table}[!ht]
\centerline{{\bf Table\,1.} Variable stars reclassified as eclipsing binaries, RS Canum Venaticorum-type}
\centerline{variables (or RS CVn), irregular variables of early spectral type, classical Cepheids}
\centerline{(or type I Cepheids), and type II Cepheids. All of our period estimates improve on}
\centerline{the GCVS periods and they are precise to the last decimal place quoted.}
\scriptsize
\begin{center}
\begin{tabular}{@{}ccccccccc@{}}
\hline
UCAC3 ID   & GCVS ID   & RA          & Dec.          & \multicolumn{2}{c}{Variable Type} & UCAC3        & \multicolumn{2}{c}{Period (d)} \\
           &           & (J2000.0)   & (J2000.0)     & GCVS         & This Work          & Aperture Mag & GCVS       & This Work         \\
\hline
169-146805 & V1018 Oph & 16 17 59.22 & $-$05 56 55.3 & RRC:         & EW                 & 15.244       & 0.3696396  & 0.350             \\
171-023140 & UY Ori    & 05 32 00.31 & $-$04 55 53.9 & RR:          & IA                 & 12.456       & -          & -                 \\
176-102611 & V0482 Hya & 08 27 38.98 & $-$02 00 34.3 & RRC:         & EW                 & 15.591       & 0.190393   & 0.3808            \\
178-131091 & V0593 Vir & 14 44 30.11 & $-$01 28 26.2 & RRC:         & EW                 & 15.473       & 0.228947   & 0.3726            \\
179-127893 & V0533 Vir & 14 12 38.56 & $-$00 53 50.7 & RRC:         & EW                 & 15.484       & 0.229537   & 0.3732            \\
180-101956 & V0491 Hya & 08 39 55.42 & $-$00 03 50.4 & RRC:         & EW                 & 14.055       & 0.263482   & 0.4170            \\
191-025421 & V0651 Ori & 05 32 46.49 & +05 24 57.8   & RR:          & EW$^{a}$           & 14.335       & -          & 0.37827           \\
192-006598 & HM Cet    & 02 07 31.42 & +05 41 05.7   & RRC          & EW$^{a}$           & 13.017       & 0.22232    & 0.44462           \\
192-026024 & V1015 Ori & 05 28 54.22 & +05 39 27.7   & RR:          & EA                 & 14.627       & -          & 1.8512            \\
205-101683 & AG Cnc    & 08 51 25.30 & +12 02 56.5   & RR:          & RS$^{b}$           & 13.684       & 0.313335   & 2.827$^{c}$       \\
208-318430 & HU Peg    & 23 59 22.17 & +13 47 11.5   & RR           & DCEP:              & 11.107       & -          & 78$^{d}$          \\
225-131002 & V0368 Her & 17 10 31.13 & +22 23 08.8   & RRAB         & DCEP:              & 15.627       & 0.543689   & 1.1915            \\
237-053427 & CN Tau    & 05 58 09.42 & +28 02 33.5   & RRAB         & DCEPS$^{e}$        & 12.645       & 0.642062   & 1.794             \\
237-121450 & UY CrB    & 16 06 21.77 & +28 07 03.8   & RR:          & CWB:$^{f}$         & 13.190       & -          & 0.92916           \\
247-041882 & DN Aur    & 05 07 59.86 & +33 23 50.7   & RRC          & EW$^{a}$           & 13.535       & 0.30846    & 0.61692           \\
263-033768 & KN Per    & 03 22 35.64 & +41 19 55.2   & RRC          & EW                 & 11.615       & 0.433224   & 0.8665            \\
268-064903 & V0421 Per & 04 45 34.83 & +43 34 22.2   & RR           & DCEP$^{g}$         & 13.643       & -          & 4.3735            \\
270-278831 & V0660 And & 23 27 52.08 & +44 54 14.9   & RRC          & EW                 & 12.147       & 0.38542    & 0.7708            \\
286-145835 & V0997 Cyg & 19 48 05.07 & +52 51 16.3   & RRC          & EW                 & 13.358       & 0.22892    & 0.45823           \\
287-146031 & V1017 Cyg & 19 56 15.81 & +53 19 12.0   & RR           & EW                 & 15.197       & 0.96       & 0.33041           \\
308-105518 & V0414 Dra & 18 53 30.15 & +63 55 03.6   & RRC:         & EW$^{a}$           & 11.306       & 0.348087   & 0.69619           \\
\hline
\end{tabular}
\raggedright \\
$^{a}$Clear detection of the O'Connell effect (i.e. unequal brightness of the two maxima). \\
$^{b}$Classification taken from van den Berg et al. (2002). \\
$^{c}$Orbital period is 2.823094~d (Mathieu et al. 2003). \\
$^{d}$This star has a single light curve in our data that spans $\sim$195 days (or $\sim$2.5 cycles). \\
$^{e}$Originally classified as a classical Cepheid by Schmidt \& Gross (1990). \\
$^{f}$Classification taken from Andrievsky et al. (2010). \\
$^{g}$Also classified as a classical Cepheid by Wils, Lloyd \& Bernhard (2006).
\end{center}
\end{table}

\IBVSfig{15cm}{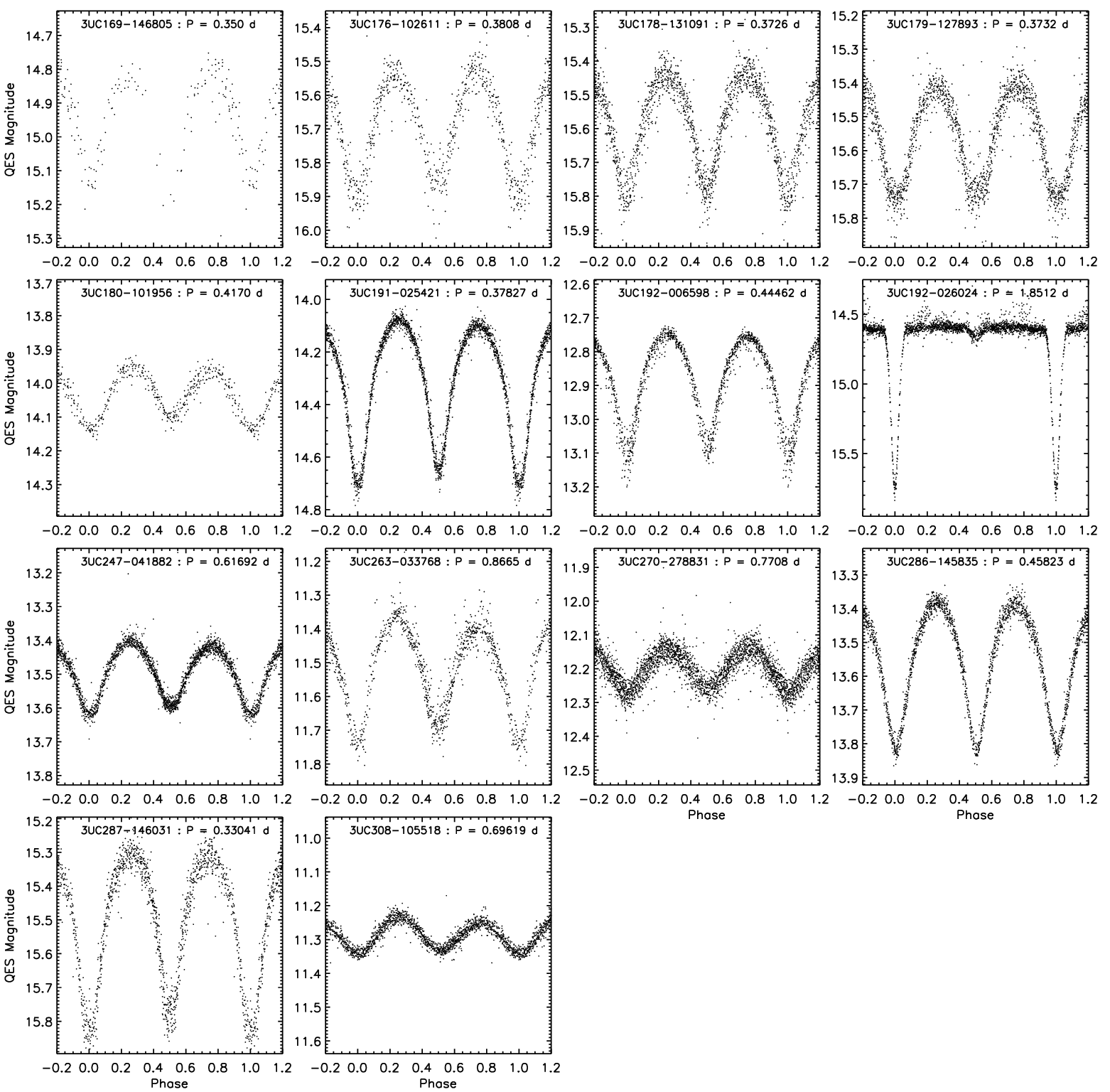}{Phased light curves of the variable stars reclassified as eclipsing binaries. The magnitude range in each plot is 0.7~mag except
                        for the stars 3UC191-025421 and 3UC192-026024 which have plots with magnitude ranges of 0.9 and 1.8~mag respectively.}
\IBVSfigKey{fig1.ps}{EB}{Phased light curves of the variable stars reclassified as eclipsing binaries.}

\IBVSfig{8cm}{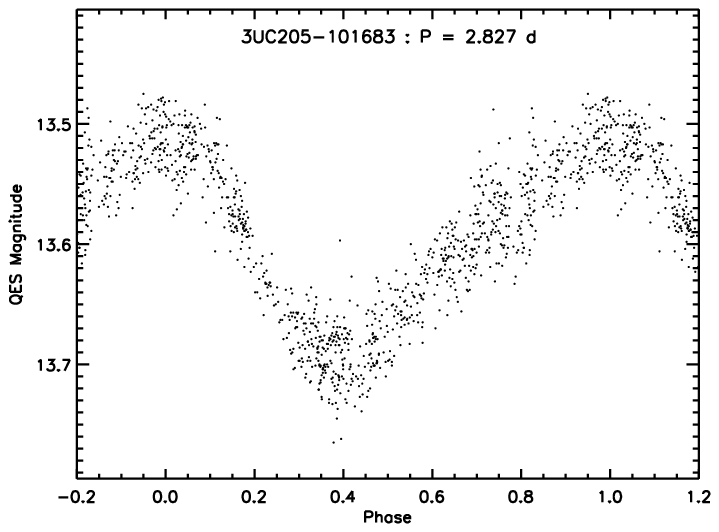}{Phased light curve of the variable star 3UC205-101683 reclassified as a RS Canum Venaticorum-type variable.}
\IBVSfigKey{fig2.ps}{RS}{Phased light curves of the variable stars reclassified as RS Canum Venaticorum-type variables.}

\IBVSfig{6.3cm}{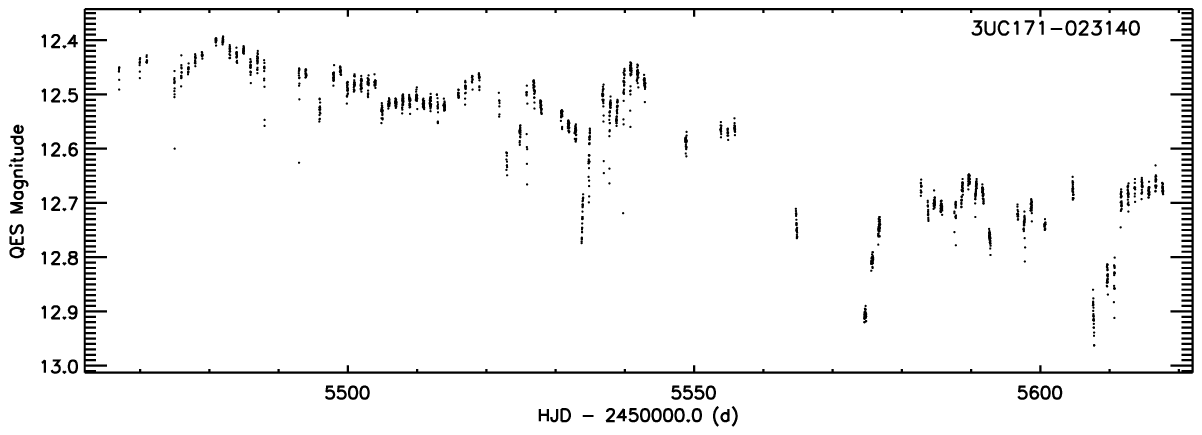}{Light curve of the variable star 3UC171-023140 reclassified as an irregular variable of early spectral type.}
\IBVSfigKey{fig3.ps}{IA}{Light curves of the variable stars reclassified as irregular variables of early spectral type.}

\IBVSfig{3.8cm}{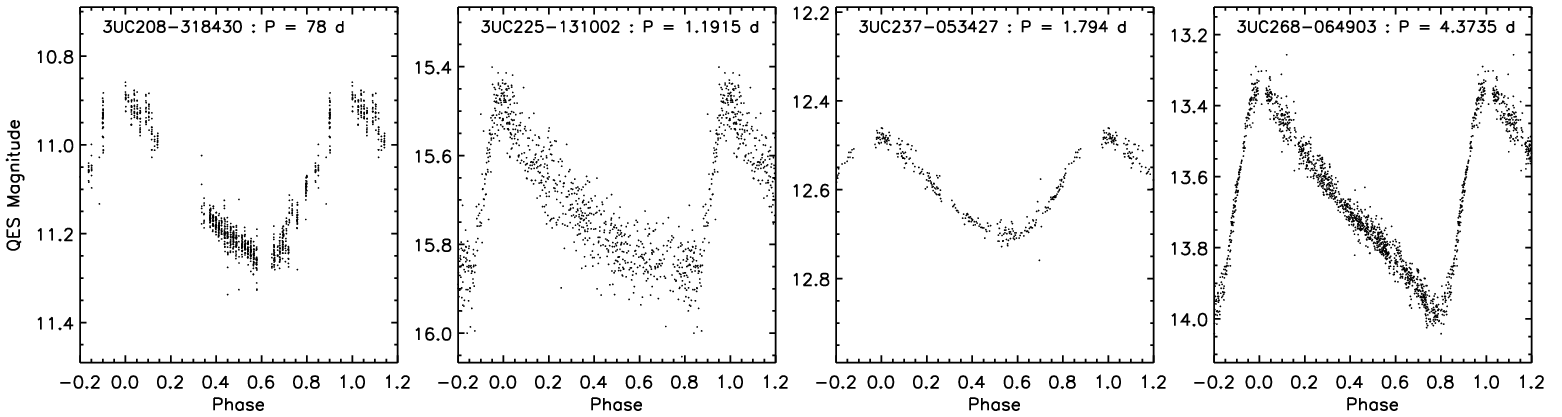}{Phased light curves of the variable stars reclassified as classical Cepheids (or type I Cepheids). The magnitude range in each plot is 0.8~mag except
                        for the star 3UC268-064903 which has a plot with a magnitude range of 1.0~mag.}
\IBVSfigKey{fig4.ps}{CEP}{Phased light curves of the variable stars reclassified as classical Cepheids.}

\IBVSfig{8cm}{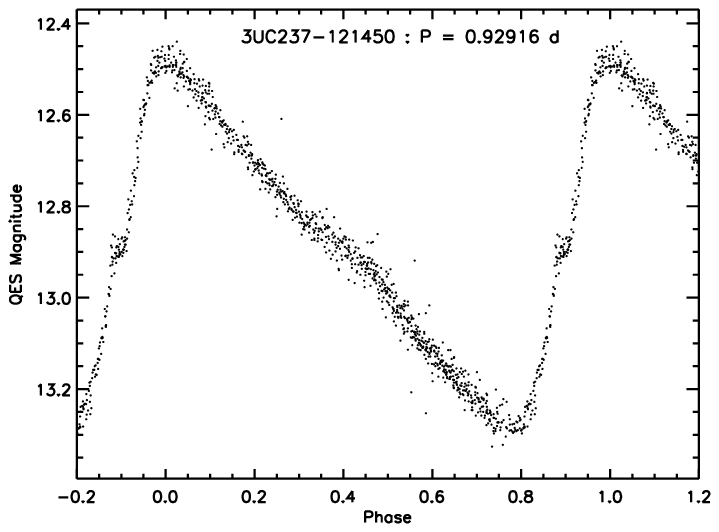}{Phased light curve of the variable star 3UC237-121450 reclassified as a type II Cepheid.}
\IBVSfigKey{fig5.ps}{CWB}{Phased light curves of the variable stars reclassified as type II Cepheids.}

\subsection{RR Lyrae stars with new sub-type classifications}

In Table~2, we report the new sub-type classifications for 61 RR Lyrae stars which have an unspecified or erroneous RR Lyrae sub-type classification
in the GCVS. We list our period estimates in the table whenever they improve on the GCVS periods (52 cases). The new sub-type classifications
are based on having considered the variable star periods and the phased light curve shapes and amplitudes. We note that some of the GCVS periods fail
to properly phase our light curves, which may indicate period changes in these cases (e.g. 3UC167-134566). We clearly detect the Blazhko effect (amplitude and/or phase modulations; Blazhko 1907)
in five of these RR Lyrae variables and this is the first detection of the effect in four of them (see the catalogue of known Galactic field Blazhko RR Lyrae stars in Skarka 2013).
For 3UC191-097728, 3UC202-143008 and 3UC227-103944, we measure
Blazhko periods of 36.2$\pm$0.2, 40.0$\pm$0.6 and 65.6$\pm$1.2~d, respectively, using the power spectrum analysis described in Section~4.5. For 3UC175-133697 we place a lower limit of 80~d on the Blazhko period.
Finally we note that the light curve for 3UC236-044458 has a strange shape for an RR Lyrae star, although its period and amplitude are consistent with that of an RRC variable.
The phase-folded light curves of all 61 variables are displayed in Figure~6.

\begin{table}[!ht]
\centerline{{\bf Table\,2.} RR Lyrae stars with new sub-type classifications. Where listed, our period}
\centerline{estimates improve on the GCVS periods and they are precise to the last decimal}
\centerline{place quoted.}
\scriptsize
\begin{center}
\begin{tabular}{@{}ccccccccc@{}}
\hline
UCAC3 ID   & GCVS ID   & RA          & Dec.          & \multicolumn{2}{c}{Variable Type} & UCAC3        & \multicolumn{2}{c}{Period (d)} \\
           &           & (J2000.0)   & (J2000.0)     & GCVS         & This Work          & Aperture Mag & GCVS       & This Work         \\
\hline
167-134566 & KW Lib    & 14 47 51.53 & $-$06 34 45.9 & RRAB         & RRC                & 14.146       & 0.3143     & 0.31269           \\
167-146876 & V0713 Oph & 16 30 10.73 & $-$06 48 02.1 & RR           & RRAB               & 14.384       & -          & 0.6858            \\
169-288913 & CH Aql    & 20 33 42.18 & $-$05 38 49.3 & RR           & RRAB               & 13.894       & 0.38918702 & -                 \\
171-261192 & V0909 Aql & 20 21 59.39 & $-$04 41 48.7 & RR:          & RRAB               & 14.758       & 0.5756     & 0.5766            \\
175-133697 & FV Lib    & 14 48 50.96 & $-$02 33 46.4 & RR:          & RRAB$^{a,b}$       & 15.028       & -          & 0.5404            \\
176-139724 & CI Ser    & 16 13 41.43 & $-$02 20 23.0 & RR           & RRAB               & 15.009       & 0.5383     & 0.5385            \\   
178-106570 & V0494 Hya & 08 43 42.26 & $-$01 20 17.0 & RRC:         & RRC                & 15.722       & 0.429765   & -                 \\
186-277530 & EL Del    & 20 55 23.08 & +02 57 35.3   & RR           & RRAB               & 14.285       & 0.595432   & -                 \\
188-017334 & FV Ori    & 04 50 43.55 & +03 47 35.1   & RR           & RRAB               & 16.486       & 0.55218    & -                 \\
188-272097 & V0911 Aql & 20 23 47.28 & +03 36 49.2   & RR           & RRAB               & 15.897       & -          & 0.46156           \\
189-092449 & V0516 Hya & 09 11 37.56 & +04 02 30.4   & RRAB:        & RRC                & 13.150       & 0.346612   & 0.34666           \\
190-098351 & UV Hya    & 09 38 15.28 & +04 45 36.5   & RR:          & RRAB               & 14.153       & -          & 0.7072            \\     
191-097728 & CY Hya    & 09 10 20.88 & +05 20 51.1   & RR           & RRAB$^{a,b}$       & 14.615       & 0.57693446 & -                 \\ 
191-135029 & V1429 Oph & 17 07 15.15 & +05 15 08.2   & RR           & RRAB               & 14.187       & -          & 0.3651            \\   
192-019411 & GO Ori    & 04 56 31.49 & +05 35 33.7   & RR           & RRAB               & 15.092       & -          & 0.53496           \\
192-101481 & IU Hya    & 09 06 17.78 & +05 45 45.3   & RR:          & RRAB               & 15.022       & -          & 0.58033           \\
192-137593 & V1053 Oph & 16 54 46.95 & +05 42 16.5   & RR:          & RRAB               & 14.921       & 4.03       & 0.578             \\   
193-136954 & V1056 Oph & 16 59 23.92 & +06 20 15.5   & RR:          & RRAB               & 16.249       & -          & 0.593             \\  
193-138339 & V2598 Oph & 17 05 43.58 & +06 25 41.5   & RRC          & RRAB               & 14.428       & 0.38749054 & 0.634             \\
194-138355 & V2620 Oph & 16 51 05.97 & +06 57 47.9   & RRAB:        & RRAB               & 15.329       & 0.456      & -                 \\
196-108163 & BF Cnc    & 08 42 12.75 & +07 48 38.0   & RR           & RRAB               & 15.928       & -          & 0.58706           \\  
196-147555 & V1600 Oph & 17 11 41.45 & +07 32 11.1   & RR           & RRC                & 15.308       & -          & 0.3080            \\   
196-147660 & V1060 Oph & 17 12 16.20 & +07 41 25.4   & RR:          & RRAB               & 15.475       & -          & 0.4404            \\    
198-016854 & CK Tau    & 04 36 44.96 & +08 54 24.4   & RR           & RRAB               & 14.808       & -          & 0.6009            \\
199-147743 & V0612 Her & 16 45 06.95 & +09 02 33.6   & RR           & RRAB               & 15.233       & -          & 0.5807            \\    
201-293397 & KL Del    & 20 38 55.50 & +10 29 03.2   & RR:          & RRAB               & 14.821       & -          & 0.44110           \\  
202-143008 & V1061 Oph & 17 14 29.98 & +10 43 07.5   & RR           & RRAB$^{a,b}$       & 14.966       & -          & 0.58940           \\
203-138413 & V1057 Oph & 17 01 06.77 & +11 03 17.6   & RR           & RRAB               & 15.398       & -          & 0.61805           \\
204-133223 & V0605 Her & 16 40 41.80 & +11 51 58.1   & RR           & RRAB               & 13.699       & -          & 0.61129           \\   
204-137112 & V1322 Oph & 17 03 43.25 & +11 51 55.5   & RR           & RRAB               & 16.071       & -          & 0.46955           \\
205-132043 & V0546 Her & 16 41 22.37 & +12 25 10.8   & RR           & RRAB               & 14.706       & -          & 0.467245          \\   
205-132516 & V0549 Her & 16 44 03.55 & +12 11 37.9   & RR           & RRAB               & 16.112       & -          & 0.58518           \\     
205-134003 & V1122 Oph & 16 53 44.99 & +12 24 46.6   & RR:          & RRAB               & 16.128       & -          & 0.50378           \\
206-140396 & V0461 Her & 17 10 49.32 & +12 52 50.9   & RR           & RRAB$^{a,b}$       & 13.264       & 0.51301    & -                 \\
207-032377 & EX Tau    & 05 44 19.60 & +13 27 54.3   & RR           & RRAB               & 15.159       & -          & 0.5556            \\
209-029649 & V0743 Ori & 05 34 58.37 & +14 25 26.9   & RR           & RRAB               & 15.722       & -          & 0.5001            \\
209-140719 & V0552 Her & 17 30 11.83 & +14 22 34.5   & RR           & RRAB               & 13.339       & -          & 0.37846           \\
215-306534 & HT Del    & 20 54 39.59 & +17 12 02.2   & RR           & RRAB               & 16.117       & 0.362494   & 0.5699            \\ 
216-136406 & BH Her    & 17 12 45.04 & +17 42 31.0   & RR:          & RRAB               & 15.793       & -          & 0.54514           \\
219-110408 & MU Boo    & 14 48 14.73 & +19 20 19.1   & RRC:         & RRC                & 14.205       & 0.320375   & -                 \\
221-099706 & GN Cnc    & 09 16 04.44 & +20 04 23.4   & RR:          & RRC                & 8.689        & -          & 0.3624            \\
222-128918 & V0383 Her & 17 16 28.23 & +20 58 44.0   & RRC          & RRAB               & 15.960       & 0.39722    & 0.56801           \\
223-112419 & CM Leo    & 11 56 14.22 & +21 15 30.2   & RRAB         & RRC                & 13.934       & 0.361732   & -                 \\
226-112429 & BU Boo    & 14 01 42.58 & +22 30 15.6   & RRAB         & RRC                & 14.853       & 0.445      & 0.4451            \\
227-103944 & AH Leo    & 11 05 05.30 & +23 21 09.0   & RR           & RRAB$^{a}$         & 14.717       & -          & 0.4663            \\
227-118042 & V0682 Her & 16 12 19.89 & +23 19 34.7   & RR           & RRC                & 15.783       & -          & 0.3102            \\   
236-044458 & IY Tau    & 05 42 23.13 & +27 56 47.6   & RRAB         & RRC$^{c}$          & 12.650       & 0.3764897  & 0.37651           \\
237-126582 & V0864 Her & 16 59 00.56 & +28 04 54.7   & RRC:         & RRC                & 15.109       & -          & 0.37537           \\
238-104806 & NW UMa    & 11 16 55.26 & +28 33 34.3   & RRAB:        & RRAB               & 15.650       & 0.5896     & 0.5895            \\
240-107056 & VZ UMa    & 11 17 28.28 & +29 40 30.1   & RR           & RRAB               & 14.452       & -          & 0.5154            \\
248-106642 & AT CVn    & 12 18 17.05 & +33 39 56.0   & RRAB:        & RRC                & 15.060       & -          & 0.3585            \\
276-118718 & BN CVn    & 12 29 36.75 & +47 49 17.3   & RR:          & RRAB               & 12.609       & -          & 0.56365           \\
279-110274 & DT UMa    & 08 53 44.85 & +49 18 40.1   & RR           & RRC                & 15.787       & -          & 0.32114           \\
282-141956 & V1104 Cyg & 19 18 00.49 & +50 45 17.8   & RR           & RRAB               & 14.797       & 0.43626    & 0.43639           \\
283-142638 & V1127 Cyg & 19 32 05.81 & +51 17 48.8   & RR           & RRAB               & 15.536       & -          & 0.64727           \\
284-142523 & V1116 Cyg & 19 24 03.28 & +51 39 52.6   & RR           & RRAB               & 15.493       & -          & 0.53853           \\
285-135326 & CD Dra    & 18 54 51.52 & +52 28 45.1   & RR           & RRAB               & 16.147       & -          & 0.5699            \\
286-140138 & V1118 Cyg & 19 24 42.97 & +52 32 50.8   & RR           & RRAB               & 15.860       & -          & 0.50654           \\  
287-136993 & V1106 Cyg & 19 19 01.50 & +53 25 15.8   & RR           & RRAB               & 15.160       & 2.04       & 0.40764           \\  
294-139890 & CI Dra    & 19 25 32.47 & +56 43 32.4   & RR           & RRAB               & 16.243       & -          & 0.47089           \\  
300-132253 & CY Dra    & 19 46 05.23 & +59 34 26.3   & RR:          & RRAB               & 12.775       & -          & 0.53494           \\
\hline
\end{tabular}
\raggedright \\
$^{a}$Clearly exhibits the Blazhko effect in our data. \\
$^{b}$Not listed in the set of known Galactic field Blazhko RR Lyrae stars from Skarka (2013). \\
$^{c}$The light curve has a strange shape for an RR Lyrae star. However, the period and amplitude are consistent with an RRC classification.
\end{center}
\end{table}

\IBVSfig{22cm}{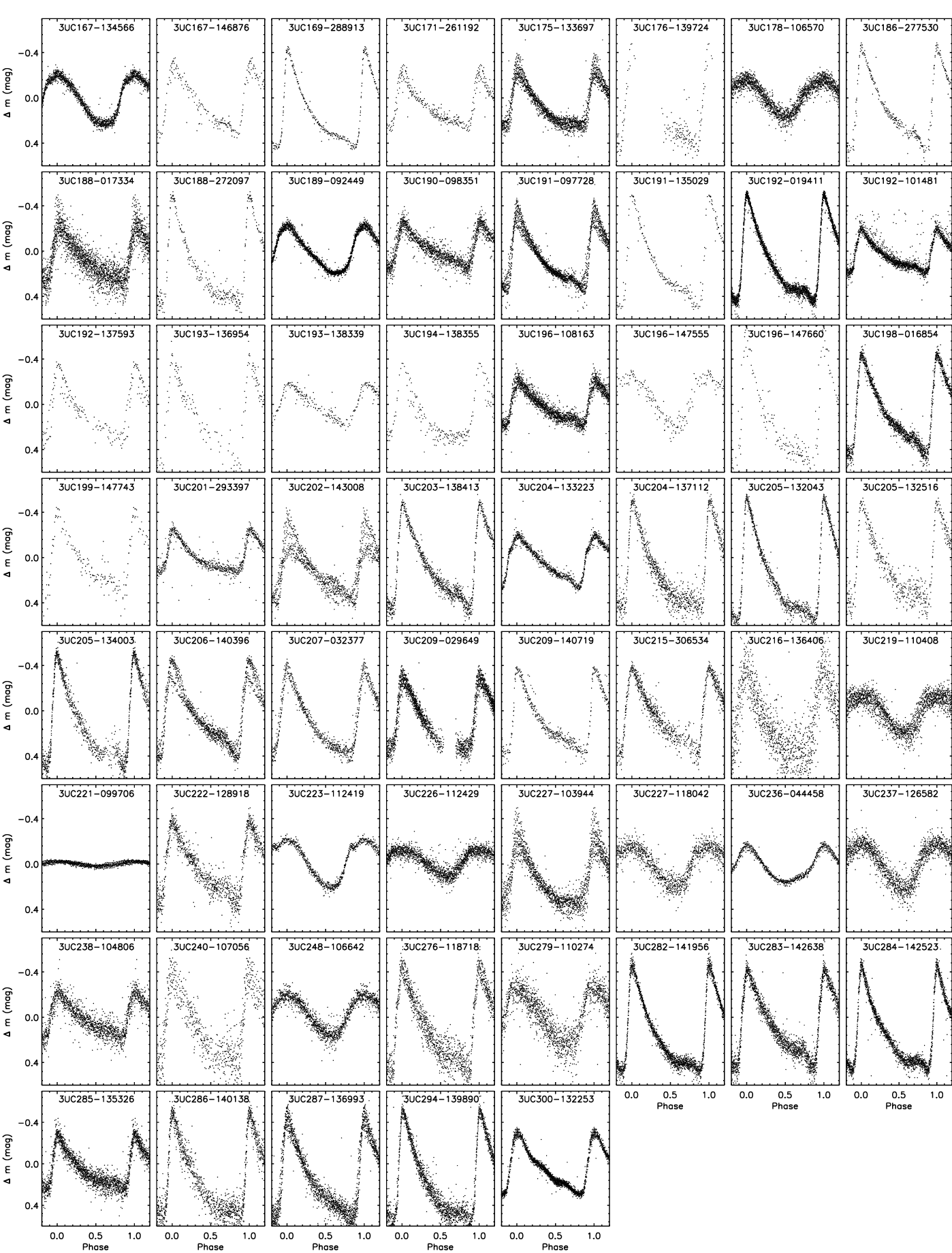}{Phased light curves of the RR Lyrae stars with new sub-type classifications. The light curve magnitudes are plotted relative to the mean
                        magnitudes and the same magnitude range of 1.3~mag is used in each plot. See Table~2 for the star brightnesses.}
\IBVSfigKey{fig6.ps}{RR}{Phased light curves of the RR Lyrae stars with new sub-type classifications.}

\subsection{Double-mode RR Lyrae stars}

Seven of the variable stars in our sample are double-mode RR Lyrae stars pulsating simultaneously in the fundamental and first overtone modes. The GCVS classification
for these stars is wrong in four cases. We checked the literature and all of these stars are known double-mode RR Lyrae stars. We used the program {\tt period04}
(Lenz \& Breger 2005) to perform a frequency analysis on the best light curve for each star. Our results are reported in Table~3 where we list the correct classification
for each star alongside the fundamental and first overtone periods that we measured. For all stars the period ratios between the two modes fall inside the expected range
of 0.742-0.748 for this type of variable star (Cox, Hodson \& Clancy 1983; Moskalik 2014). The corresponding light curves phased using the first overtone period are plotted in Figure~7.

\subsection{RR Lyrae stars with improved periods}

For the remaining 499 variables in our sample, the RR Lyrae classifications in the GCVS are correct. However, we have been able to improve
on the GCVS period estimates for 83 variables. These stars along with their improved periods are listed in Table~4. Again, period changes in some of
these variables may explain the differences between the GCVS and our period estimates (e.g. 3UC204-103494). Note that the GCVS period estimates
are the best periods available for the other 416 RR Lyrae variables in our sample. We clearly detect the Blazhko effect in ten of
the variables listed in Table~4 and this is the first detection of the effect in seven of them (see Skarka 2013). For 3UC188-089887, 3UC192-101314, 3UC209-135992 and 3UC234-000057,
we measure Blazhko periods of 69.7$\pm$0.5, 56.6$\pm$0.7, 38.8$\pm$0.2 and 80$\pm$4~d, respectively, using the power spectrum analysis described in Section~4.5. For 3UC232-112040 and
3UC282-145093 we place lower limits of 100 and 80~d, respectively, on the Blazhko periods.

\subsection{New detections of the Blazhko effect in RR Lyrae stars}

Finally, while inspecting the light curves of the remaining 499 RR Lyrae stars in our sample, we looked for clear indications of amplitude and phase modulations that characterise the Blazhko effect.
We then checked our suspected Blazhko RR Lyrae stars against the catalogue of known Galactic field Blazhko stars in Skarka (2013). We found 27 RR Lyrae stars which clearly exhibit the Blazhko effect and which
are not in the Skarka (2013) catalogue. This brings the total number of RR Lyrae stars where we have detected the Blazhko effect for the first time to 38 when taking into account the
4 and 7 such stars listed in Tables~2 and~4, respectively. We confirmed the presence of the Blazhko effect in 19 of these 27 variables by analysing the power spectra of the
light curves using {\tt period04}. We did this by prewhitening the power spectrum for the primary frequency $f_{0}$ (and the harmonics where necessary)
and considered the Blazhko effect to have been detected in the power spectrum if the next highest peak $f_{\mbox{\scriptsize peak}}$ has a significant amplitude ($>$0.02-0.05 mag depending on light curve quality) and
a ratio to the primary frequency in the range $\sim$0.95-1.05 (Benk{\H o}, Szab\'o \& Papar\'o 2011).
The Blazhko period is then estimated via $P_{\mbox{\scriptsize bl}} =  1 / \left| f_{\mbox{\scriptsize peak}} - f_{0} \right |$.
We present the details of these Blazhko variables in Table~5 and we plot the phased light curves in Figure~8. In four cases where we could not estimate the Blazhko period
from the power spectrum, we were still able to place lower limits on the Blazhko period by inspecting the unphased light curve.

\subsection{Electronic light curve data}

We provide the 1783 light curves for the sample of 588 photometrically variable objects described in this paper in an electronic table. The 588
variables breakdown by type as follows: 482 RRAB, 78 RRC, 7 RR(B), 13 EW, 1 EA, 1 RS, 1 IA, 3 DCEP, 1 DCEPS and 1 CWB. An excerpt from the electronic
table is presented in Table~6. The light curves will also be made available via CDS (Strasbourg) where we hope that the data will be of
further use to the astronomical community.

\bigskip

ACKNOWLEDGMENTS: This research made use of the SIMBAD database, the VizieR catalogue access tool, and the
cross-match service provided by CDS, Strasbourg, France. This publication was made possible by NPRP
grant \# X-019-1-006 from the Qatar National Research Fund (a member of Qatar Foundation). The statements
made herein are solely the responsibility of the authors.
AAF is grateful to DGAPA-UNAM for grant number IN104612.
Thanks goes to Noe Kains at STScI in Baltimore
for hosting the first author during part of this work and for the many useful discussions.

\references

Alard C.; \& Lupton R.H. 1998, {\it ApJ}, {\bf 503}, 325

Alsubai K.A.; Parley N.R.; Bramich D.M.; Horne K.; Collier Cameron A.; West R.G.; Sorensen P.M.; Pollacco D.; Smith J.C.; \& Fors O. 2013, {\it Acta Astronomica}, {\bf 63}, 465

Alsubai K.A. et al. 2011, {\it MNRAS}, {\bf 417}, 709

Andrievsky S.M.; Kovtyukh V.V.; Wallerstein G.; Korotin S.A.; \& Huang W., {\it PASP}, {\bf 122}, 877

Belloni T.; Verbunt F.; \& Mathieu R.D. 1998, {\it A\&A}, {\bf 339}, 431

Benk{\H o} J.M.; Szab\'o R.; \& Papar\'o M. 2011, {\it MNRAS}, {\bf 417}, 975

Bernhard K. 2010, {\it BAV Rundbrief 2/2010}, {\bf 59}, 78

Blazhko S. 1907, {\it Astronomische Nachrichten}, {\bf 175}, 325

Bramich D.M.; Horne K.; Albrow M.D.; Tsapras Y.; Snodgrass C.; Street R.A.; Hundertmark M.; Kains N.; Arellano Ferro A.; Figuera Jaimes R.; \& Giridhar S. 2013, {\it MNRAS}, {\bf 428}, 2275

Bramich D.M. 2008, {\it MNRAS}, {\bf 386}, 77

Bryan M.L. et al. 2012, {\it ApJ}, {\bf 750}, 84

Burke E.W.; Rolland W.W.; \& Boy W.R. 1970, {\it Journal of the Royal Astronomical Society of Canada}, {\bf 64}, 353

Cox A.N.; Hodson S.W.; \& Clancy S.P. 1983, {\it ApJ}, {\bf 266}, 94

Dworetsky M.M. 1983, {\it MNRAS}, {\bf 203}, 917

Gomez-Forrellad J.M.; \& Garcia-Melendo E. 1995, {\it IBVS}, No. 4247

Lenz P.; \& Breger M. 2005, {\it Communications in Asteroseismology}, {\bf 146}, 53
           
Mathieu R.D.; van den Berg M.; Torres G.; Latham D.; Verbunt F.; \& Stassun K. 2003, {\it AJ}, {\bf 125}, 246

Moskalik P. 2014, {\it Proceedings of the International Astronomical Union}, {\bf 301}, 249

O'Connell D.J.K. 1951, {\it Riverview College Observatory Publications}, {\bf 2}, 85

Samus N.N. et al. 2009, {\it General Catalogue of Variables Stars (Vol. I-III, version 30/04/2013)}, {\bf 2009yCat....102025S}

Schmidt E.G. \& Gross B.A. 1990, {\it PASP}, {\bf 102}, 978

Skarka M. 2013, {\it A\&A}, {\bf 549}, 101

van den Berg M.; Stassun K.G.; Verbunt F.; \& Mathieu R.D. 2002, {\it A\&A}, {\bf 382}, 888

Vieira S.L.A. et al. 2003, {\it AJ}, {\bf 126}, 2971

Wallerstein G.; Kovtyukh V.V.; \& Andrievsky S.M. 2009, {\it ApJ}, {\bf 692}, 127

Wils P.; Lloyd C.; \& Bernhard K. 2006, {\it MNRAS}, {\bf 368}, 1757
              
Zacharias N. et al. 2010, {\it AJ}, {\bf 139}, 2184

\endreferences

\begin{table}[!ht]
\centerline{{\bf Table\,3.} Double-mode RR Lyrae stars in our sample that are pulsating simultaneously in}
\centerline{the fundamental and first overtone modes. In column~9, we list the fundamental and first}
\centerline{overtone periods $P_{0}$ and $P_{1}$, respectively, along with the period ratio $P_{1} / P_{0}$ that we}
\centerline{measure from our data. These periods are precise to the last decimal place quoted.}
\scriptsize
\begin{center}
\begin{tabular}{@{}ccccccccc@{}}
\hline
UCAC3 ID   & GCVS ID   & RA          & Dec.          & \multicolumn{2}{c}{Variable Type} & UCAC3        & Period (d) & Period (d)                        \\
           &           & (J2000.0)   & (J2000.0)     & GCVS         & This Work          & Ap. Mag      & GCVS       & This Work : $P_{0}$, $P_{1}$, $P_{1} / P_{0}$ \\
\hline
173-108416 & V0500 Hya & 08 47 46.93 & $-$03 39 00.3 & RR(B)        & RR(B)              & 10.661       & 0.42079    & 0.5639, 0.4208, 0.7462                   \\
183-255136 & QW Aqr    & 21 07 26.08 & +01 10 17.6   & RR(B)        & RR(B)              & 13.794       & 0.35498    & 0.4772, 0.3551, 0.7441                   \\
201-119785 & AQ Leo    & 11 23 55.28 & +10 18 59.1   & RR(B)        & RR(B)              & 12.679       & 0.5497508  & 0.5498, 0.4102, 0.7461                   \\
206-276118 & CF Del    & 20 23 31.36 & +12 59 30.5   & RR           & RR(B)              & 14.307       & 0.49923    & 0.47843, 0.35604, 0.74418               \\
218-133356 & V0458 Her & 17 08 30.92 & +18 31 14.3   & RRC          & RR(B)              & 13.305       & 0.3599801  & 0.48352, 0.35998, 0.74450                \\
248-100379 & WY LMi    & 09 30 23.25 & +33 53 10.6   & RRAB         & RR(B)              & 15.391       & 0.420003   & 0.4923, 0.3662, 0.7439                   \\
263-117859 & BN UMa    & 11 16 22.91 & +41 14 01.4   & RRC          & RR(B)              & 13.787       & 0.399901   & 0.53594, 0.39965, 0.74571                \\
\hline
\end{tabular}
\end{center}
\end{table}

\IBVSfig{7.5cm}{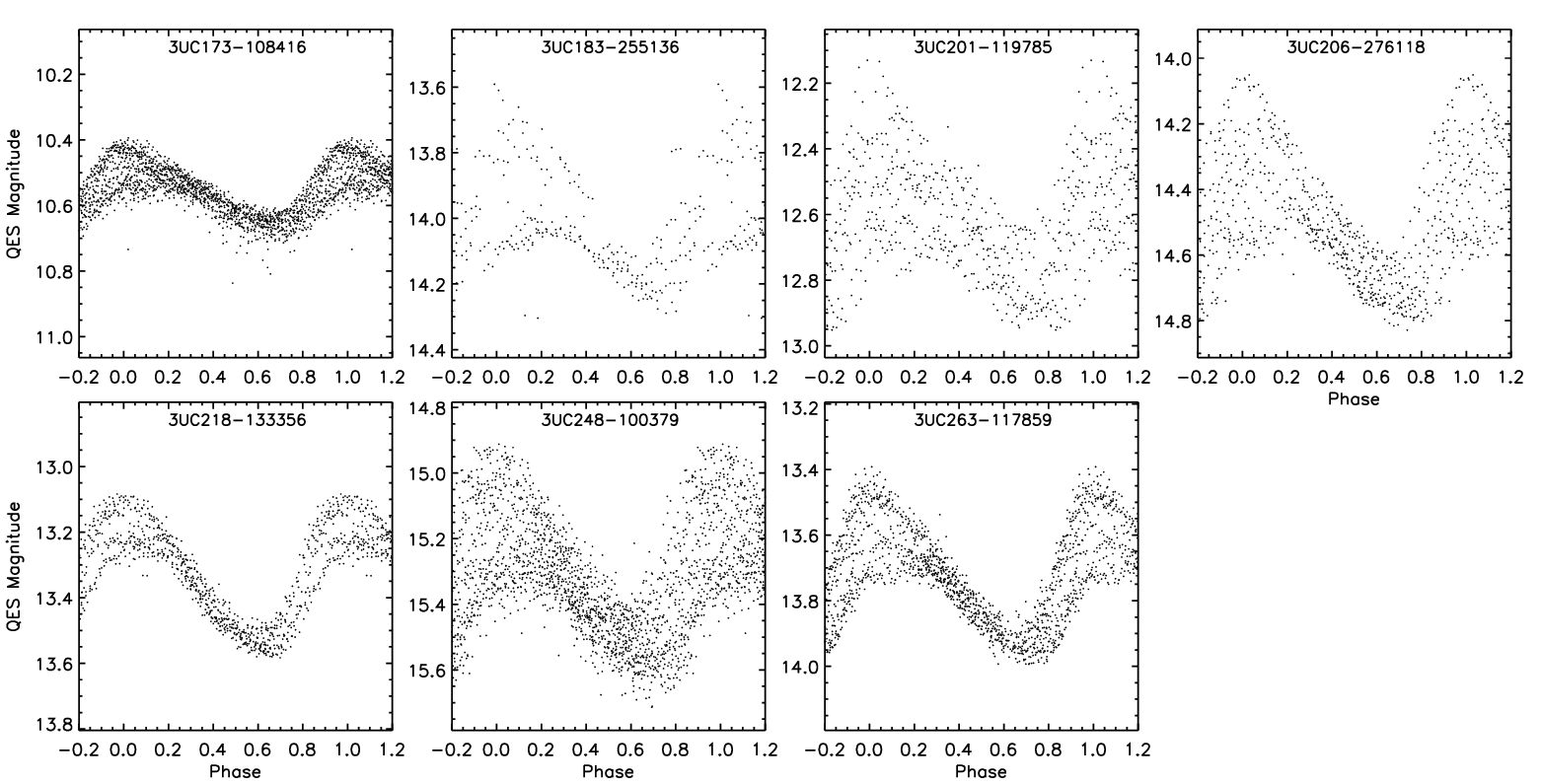}{Phased light curves of the double-mode RR Lyrae stars. The light curves are phased with the first overtone period. The magnitude
                       range in each plot is 1.0~mag.}
\IBVSfigKey{fig7.ps}{RR(B)}{Phased light curves of the double-mode RR Lyrae stars.}

\begin{table}[!ht]
\centerline{{\bf Table\,4.} RR Lyrae stars with improved periods. Our periods are precise to the last}
\centerline{decimal place quoted.}
\scriptsize
\begin{center}
\begin{tabular}{@{}cccccccc@{}}
\hline
UCAC3 ID   & GCVS ID   & RA          & Dec.          & Variable Type & UCAC3        & \multicolumn{2}{c}{Period (d)} \\
           &           & (J2000.0)   & (J2000.0)     &               & Aperture Mag & GCVS       & This Work         \\
\hline
170-106909 & DG Hya    & 08 58 06.36 & $-$05 26 25.2 & RRAB          & 12.569       & 0.429973   & 0.75425           \\
173-130413 & HR Vir    & 13 42 28.63 & $-$03 37 32.7 & RRAB          & 14.414       & -          & 0.7394            \\ 
176-019652 & V0964 Ori & 05 07 54.52 & $-$02 08 48.7 & RRAB          & 13.267       & 0.5046561  & 0.50464           \\
177-017006 & V1830 Ori & 04 49 34.97 & $-$01 42 19.5 & RRC           & 15.955       & 0.276438   & 0.2734            \\
177-261611 & V0910 Aql & 20 23 11.69 & $-$01 33 57.5 & RRAB          & 14.725       & 1.0        & 0.50019           \\
179-018625 & V1844 Ori & 05 03 36.84 & $-$00 59 57.1 & RRAB          & 15.057       & 0.778216   & 0.58908           \\
179-140408 & V0694 Oph & 16 22 47.53 & $-$00 49 37.5 & RRAB          & 14.845       & 0.62       & 0.6207            \\
179-141655 & V0714 Oph & 16 30 03.08 & $-$00 59 56.5 & RRAB          & 14.543       & 0.556      & 0.5557            \\
182-001386 & BF Cet    & 00 27 03.97 & +00 40 30.3   & RRC           & 13.856       & -          & 0.38034           \\
188-089887 & CW Hya    & 08 55 07.81 & +03 39 24.7   & RRAB$^{a,b}$  & 15.904       & 0.4820734  & 0.48050           \\
192-101314 & V0430 Hya & 09 04 48.58 & +05 30 08.3   & RRAB$^{a,b}$  & 12.803       & 0.49691    & 0.496830          \\
193-135481 & V2509 Oph & 16 51 29.85 & +06 22 26.2   & RRAB          & 13.352       & -          & 0.7786            \\
194-125256 & GT Vir    & 14 56 48.38 & +06 48 27.7   & RRAB          & 15.089       & 0.4080564  & 0.68931           \\  
195-285152 & LX Del    & 20 52 18.98 & +07 08 46.8   & RRAB          & 13.876       & -          & 0.5669            \\
197-007736 & BP Cet    & 02 24 52.15 & +08 24 05.0   & RRAB          & 14.897       & -          & 0.6924            \\
197-143679 & V1013 Her & 16 24 49.66 & +08 04 14.1   & RRAB          & 13.258       & -          & 0.6448            \\
199-293401 & LW Del    & 20 38 27.40 & +09 12 05.4   & RRAB          & 12.890       & -          & 0.5811            \\
200-000971 & FF Psc    & 00 17 48.58 & +09 53 22.1   & RRAB          & 12.441       & 0.70119    & 0.70110           \\
201-029115 & V0944 Ori & 05 36 11.40 & +10 29 23.0   & RRAB          & 15.456       & -          & 0.5873            \\
201-114082 & DL Leo    & 09 43 03.58 & +10 19 01.3   & RRAB          & 13.551       & -          & 0.67378           \\
203-295430 & DG Del    & 20 35 44.20 & +11 28 09.0   & RRAB          & 15.797       & 0.326961   & 0.4905            \\
203-295942 & DI Del    & 20 36 49.52 & +11 20 21.9   & RRAB          & 15.487       & 0.367221   & 0.5803            \\
204-103494 & AM Cnc    & 08 56 14.84 & +11 37 20.6   & RRAB          & 14.815       & 0.557615   & 0.55803           \\
204-113713 & GP Leo    & 11 45 45.53 & +11 52 08.5   & RRAB          & 13.952       & -          & 0.6793            \\
204-293402 & HV Del    & 20 33 19.53 & +11 32 01.7   & RRAB          & 15.636       & 0.721265   & 0.5649            \\
207-115598 & LL Leo    & 11 30 53.62 & +13 19 28.4   & RRAB          & 13.144       & 0.3324     & 0.33239           \\
209-135992 & V1124 Her & 17 04 32.86 & +14 26 32.7   & RRAB$^{a}$    & 12.426       & 0.551      & 0.55103           \\
211-103053 & KW Cnc    & 08 40 47.96 & +15 24 52.4   & RRAB          & 14.591       & 0.60102    & 0.60044           \\
211-128841 & AW Ser    & 16 06 28.79 & +15 22 05.8   & RRAB          & 12.983       & 0.597097   & 0.597114          \\
211-291434 & CS Del    & 20 28 54.87 & +15 13 14.0   & RRC           & 12.912       & 0.365737   & 0.37088           \\
212-310112 & V0398 Peg & 21 08 57.95 & +15 56 55.3   & RRAB          & 13.893       & 0.55259    & 0.55136           \\
213-110778 & HY Com    & 12 18 16.02 & +16 09 15.9   & RRC           & 10.281       & -          & 0.4488            \\
214-112667 & BX Leo    & 11 38 02.06 & +16 32 36.2   & RRC           & 11.771       & 0.36286    & 0.36277           \\
215-288176 & CH Del    & 20 23 18.39 & +17 06 13.5   & RRC           & 13.176       & 0.4596     & 0.31499           \\
216-128398 & V0686 Her & 16 14 23.25 & +17 56 35.2   & RRAB          & 14.800       & 0.510987   & 0.511004          \\
216-129723 & V0695 Her & 16 25 58.65 & +17 42 52.0   & RRAB$^{a,b}$  & 14.574       & 0.678788   & 0.67884           \\
216-339177 & V0611 Peg & 23 46 41.17 & +17 38 02.6   & RRAB          & 13.794       & 0.58868    & 0.588665          \\
216-339421 & V0419 Peg & 23 50 05.03 & +17 53 44.0   & RRAB          & 14.674       & 0.60373    & 0.60370           \\
218-287254 & EO Del    & 20 37 47.72 & +18 55 30.6   & RRAB          & 14.378       & 0.580861   & 0.57990           \\
219-270870 & II Del    & 20 50 01.18 & +19 11 43.5   & RRC           & 14.558       & 0.408021   & 0.4078            \\
224-130750 & SW Her    & 16 58 27.56 & +21 32 51.5   & RRAB          & 14.956       & 0.49287277 & 0.492861          \\
225-123239 & V0504 Ser & 16 01 52.29 & +22 22 47.9   & RRAB          & 14.169       & 0.56396    & 0.563833          \\
225-131962 & V0382 Her & 17 16 17.21 & +22 01 04.5   & RRAB          & 15.934       & 0.4556118  & 0.45554           \\
225-265413 & FH Vul    & 20 40 19.89 & +22 13 24.3   & RRAB          & 13.267       & 0.4054185  & 0.405412          \\
227-031814 & HX Tau    & 05 10 48.79 & +23 12 22.7   & RRAB          & 15.298       & 0.53875    & 0.53826           \\
227-119598 & V0362 Her & 16 30 39.55 & +23 26 41.7   & RRAB          & 14.791       & 0.718297   & 0.7185            \\
228-098638 & EZ Cnc    & 08 52 57.67 & +23 47 54.2   & RRAB          & 12.404       & -          & 0.54578           \\
228-261882 & BL Peg    & 21 22 59.51 & +23 53 32.1   & RRAB$^{a,b}$  & 14.433       & 0.55543    & 0.55555           \\
229-254073 & V0507 Vul & 20 49 45.85 & +24 12 44.9   & RRC           & 11.848       & 0.336126   & 0.33607           \\
230-118556 & V0677 Her & 16 08 04.15 & +24 59 20.2   & RRAB          & 14.387       & 0.475716   & 0.475728          \\
230-120541 & V1186 Her & 16 29 14.78 & +24 59 38.7   & RRAB          & 13.894       & 0.44032    & 0.44025           \\
231-126743 & V0467 Her & 17 12 50.79 & +25 01 48.6   & RRAB          & 14.850       & 0.6835066  & 0.65352           \\
232-094191 & AS Cnc    & 08 25 42.16 & +25 43 08.5   & RRAB          & 12.998       & 0.61752    & 0.61754           \\
232-096427 & SX Cnc    & 08 51 19.58 & +25 33 24.3   & RRAB          & 14.026       & 0.5101754  & 0.51016           \\
232-111124 & IY Boo    & 14 19 39.22 & +25 47 24.1   & RRAB          & 14.462       & 0.59165    & 0.59171           \\
232-112040 & LN Boo    & 14 37 09.05 & +25 44 46.6   & RRAB$^{a,b}$  & 13.997       & 0.46675    & 0.46667           \\
234-000057 & GV Peg    & 00 00 35.59 & +26 39 49.5   & RRAB$^{a,b}$  & 13.600       & 0.5669237  & 0.56607           \\
235-118691 & CT CrB    & 16 18 34.34 & +27 28 13.2   & RRAB$^{a}$    & 14.271       & 0.508646   & 0.50858           \\
236-097672 & KV Cnc    & 08 40 02.43 & +27 43 31.5   & RRAB$^{a}$    & 12.916       & 0.502      & 0.5020            \\
236-123949 & V0860 Her & 16 50 38.71 & +27 58 40.3   & RRAB          & 16.051       & -          & 0.57083           \\
237-108110 & EF Leo    & 11 49 10.95 & +28 00 25.6   & RRAB          & 15.891       & -          & 0.5979            \\
237-274829 & V0466 Vul & 21 05 22.87 & +28 17 49.4   & RRAB          & 14.752       & 0.4759     & 0.47592           \\
239-000617 & IQ Peg    & 00 06 05.70 & +29 19 12.6   & RRAB          & 16.319       & -          & 0.47993           \\
240-120704 & RV CrB    & 16 19 25.85 & +29 42 47.6   & RRC           & 11.387       & 0.331565   & 0.33172           \\
245-013999 & VX Tri    & 02 10 00.87 & +32 24 08.9   & RRAB          & 14.576       & -          & 0.6331            \\
247-105751 & CK Com    & 12 14 50.60 & +33 06 05.9   & RRAB          & 14.727       & 0.6939962  & 0.6938            \\
248-104694 & V0345 UMa & 11 17 49.43 & +33 40 14.8   & RRAB          & 14.434       & 0.57667    & 0.57662           \\
248-106344 & DN CVn    & 12 09 17.00 & +33 39 35.5   & RRC           & 15.083       & 0.3266873  & 0.32625           \\
251-289738 & DM And    & 23 32 00.72 & +35 11 48.9   & RRAB          & 11.978       & 0.630387   & 0.6305            \\
254-099280 & VY LMi    & 09 27 41.32 & +36 58 22.4   & RRAB          & 13.838       & -          & 0.52614           \\
255-095249 & DQ Lyn    & 08 23 40.99 & +37 28 10.8   & RRC           & 11.670       & -          & 0.4949            \\
257-098638 & EN Lyn    & 08 46 07.04 & +38 02 52.7   & RRAB          & 13.554       & 0.6249     & 0.6251            \\
262-118704 & AO UMa    & 11 07 39.83 & +40 33 57.2   & RRAB          & 15.749       & 0.561614   & 0.56265           \\
263-257028 & DY And    & 23 58 42.21 & +41 29 19.4   & RRAB          & 13.674       & 0.603087   & 0.60323           \\
266-126680 & AQ UMa    & 11 12 59.53 & +42 48 41.7   & RRAB          & 16.369       & 0.644029   & 0.6433            \\
266-127186 & AV UMa    & 11 29 40.53 & +42 44 24.7   & RRAB          & 15.935       & 0.479483   & 0.47911           \\
272-115635 & AX UMa    & 11 38 26.84 & +45 56 05.9   & RRAB          & 13.591       & 0.53491    & 0.53468           \\
278-047793 & AN Per    & 03 08 31.34 & +48 32 40.4   & RRAB          & 14.367       & 0.602818   & 0.6021            \\
282-136249 & DT Dra    & 18 49 57.25 & +50 35 12.8   & RRAB          & 13.625       & -          & 0.52664           \\
282-145093 & V1949 Cyg & 19 30 12.47 & +50 48 21.2   & RRAB$^{a,b}$  & 13.714       & -          & 0.4994            \\
293-139097 & XZ Cyg    & 19 32 29.31 & +56 23 17.5   & RRAB          & 9.500        & 0.4667     & 0.4666            \\
297-143410 & V1035 Cyg & 20 05 41.44 & +58 02 48.9   & RRAB          & 15.798       & 0.5321     & 0.5318            \\
312-073056 & V0420 Dra & 19 18 09.26 & +65 35 17.7   & RRC           & 12.780       & 0.32963    & 0.32951           \\
\hline
\end{tabular}
\raggedright \\
$^{a}$Clearly exhibits the Blazhko effect in our data. \\
$^{b}$Not listed in the set of known Galactic field Blazhko RR Lyrae stars from Skarka (2013).
\end{center}
\end{table}

\begin{table}[!ht]
\centerline{{\bf Table\,5.} New detections of the Blazhko effect in Galactic RR Lyrae stars. These stars}
\centerline{are not listed in the catalogue of Skarka (2013).}
\scriptsize
\begin{center}
\begin{tabular}{@{}cccccccc@{}}
\hline
UCAC3 ID   & GCVS ID   & RA          & Dec.          & Variable  & UCAC3        & Period (d) & Blazhko       \\
           &           & (J2000.0)   & (J2000.0)     & Type      & Aperture Mag & GCVS       & Period (d)    \\
\hline
162-283238 & PQ Aqr    & 20 43 15.75 & $-$09 09 28.7 & RRAB      & 13.713       & 0.512286   & -             \\ 
164-130527 & V0574 Vir & 14 34 30.48 & $-$08 18 32.7 & RRAB      & 14.529       & 0.47439    & 26.3$\pm$0.3  \\
172-132806 & V0586 Vir & 14 39 47.52 & $-$04 08 05.3 & RRAB      & 13.376       & 0.682772   & 132$\pm$3     \\
174-136150 & V0585 Vir & 14 39 27.36 & $-$03 27 36.6 & RRAB      & 12.872       & 0.601615   & 93.8$\pm$0.4  \\
174-247961 & PS Aqr    & 20 44 03.69 & $-$03 23 12.3 & RRAB      & 14.411       & 0.59102    & -             \\
175-104590 & V0486 Hya & 08 30 29.83 & $-$02 42 36.8 & RRAB      & 13.107       & 0.508655   & 18.5$\pm$1.0  \\
176-131910 & MS Lib    & 14 53 32.99 & $-$02 06 51.5 & RRAB      & 14.445       & 0.441448   & 105$\pm$10    \\
178-129959 & V0561 Vir & 14 28 40.65 & $-$01 27 59.4 & RRAB      & 15.893       & 0.550276   & 42.0$\pm$0.6  \\
186-091699 & V0487 Hya & 08 32 57.03 & +02 59 02.9   & RRAB      & 13.447       & 0.561485   & 64.4$\pm$0.3  \\
186-092895 & GL Hya    & 08 40 59.22 & +02 37 22.2   & RRAB      & 13.409       & 0.50593    & 157$\pm$10    \\
193-094864 & V0425 Hya & 08 20 51.78 & +06 28 24.2   & RRAB      & 14.829       & 0.5508     & 61.1$\pm$0.2  \\
194-123163 & AF Vir    & 14 28 09.82 & +06 32 43.9   & RRAB      & 11.507       & 0.48376    & 35$\pm$5      \\
201-137842 & V1162 Her & 16 17 00.49 & +10 17 27.9   & RRAB      & 13.429       & 0.547925   & -             \\
204-017429 & V1327 Tau & 04 40 09.89 & +11 43 17.0   & RRC       & 13.169       & 0.3312     & 23.7$\pm$0.3  \\
206-121944 & UY Boo    & 13 58 46.34 & +12 57 06.5   & RRAB      & 11.001       & 0.6508964  & -             \\
208-001501 & FI Psc    & 00 23 22.80 & +13 45 40.8   & RRAB      & 13.590       & 0.53129    & $>$120        \\
209-121247 & LS Boo    & 14 38 21.77 & +14 24 55.1   & RRAB      & 13.624       & 0.5527108  & 42.9$\pm$1.9  \\
216-112989 & AE Leo    & 11 26 12.19 & +17 39 39.7   & RRAB      & 12.508       & 0.626723   & 62$\pm$4      \\
216-121383 & LW Boo    & 14 40 32.61 & +17 35 57.3   & RRAB      & 13.434       & 0.56342    & 63.9$\pm$0.6  \\
217-331820 & V0606 Peg & 23 41 58.75 & +18 13 01.4   & RRAB      & 12.564       & 0.52966    & 26.7$\pm$0.3  \\
228-114655 & DD Boo    & 14 51 20.08 & +23 32 30.0   & RRC$^{a}$ & 12.889       & 0.3393397  & 9.64$\pm$0.02 \\
237-128729 & V0385 Her & 17 16 26.66 & +28 05 56.4   & RRAB      & 15.499       & 0.5281428  & $>$100        \\
238-035854 & NU Aur    & 05 09 02.37 & +28 40 52.7   & RRAB      & 13.436       & 0.53941672 & $>$60         \\
239-115076 & XX Boo    & 14 51 37.56 & +29 21 26.7   & RRAB      & 12.061       & 0.581402   & 148$\pm$7     \\
248-099777 & FW Lyn    & 09 19 51.49 & +33 52 23.9   & RRAB      & 13.650       & 0.52174    & $>$110        \\
296-130079 & NQ Dra    & 18 44 13.17 & +57 40 59.4   & RRAB      & 13.752       & 0.52919    & 34.5$\pm$0.4  \\
304-116701 & V0429 Dra & 19 59 32.15 & +61 31 21.0   & RRAB      & 14.994       & 0.5862     & 87.0$\pm$2.2  \\
\hline
\end{tabular}
\raggedright \\
$^{a}$Although the amplitude modulations are small, they are clear to the eye in the unphased light curve and they are strongly
detected in the power spectrum. To help confirm our conclusions for this star, we checked the literature and found that
Gomez-Forrellad \& Garcia-Melendo (1995) also suspected this star of exhibiting the Blazhko effect.
\end{center}
\end{table}

\IBVSfig{22cm}{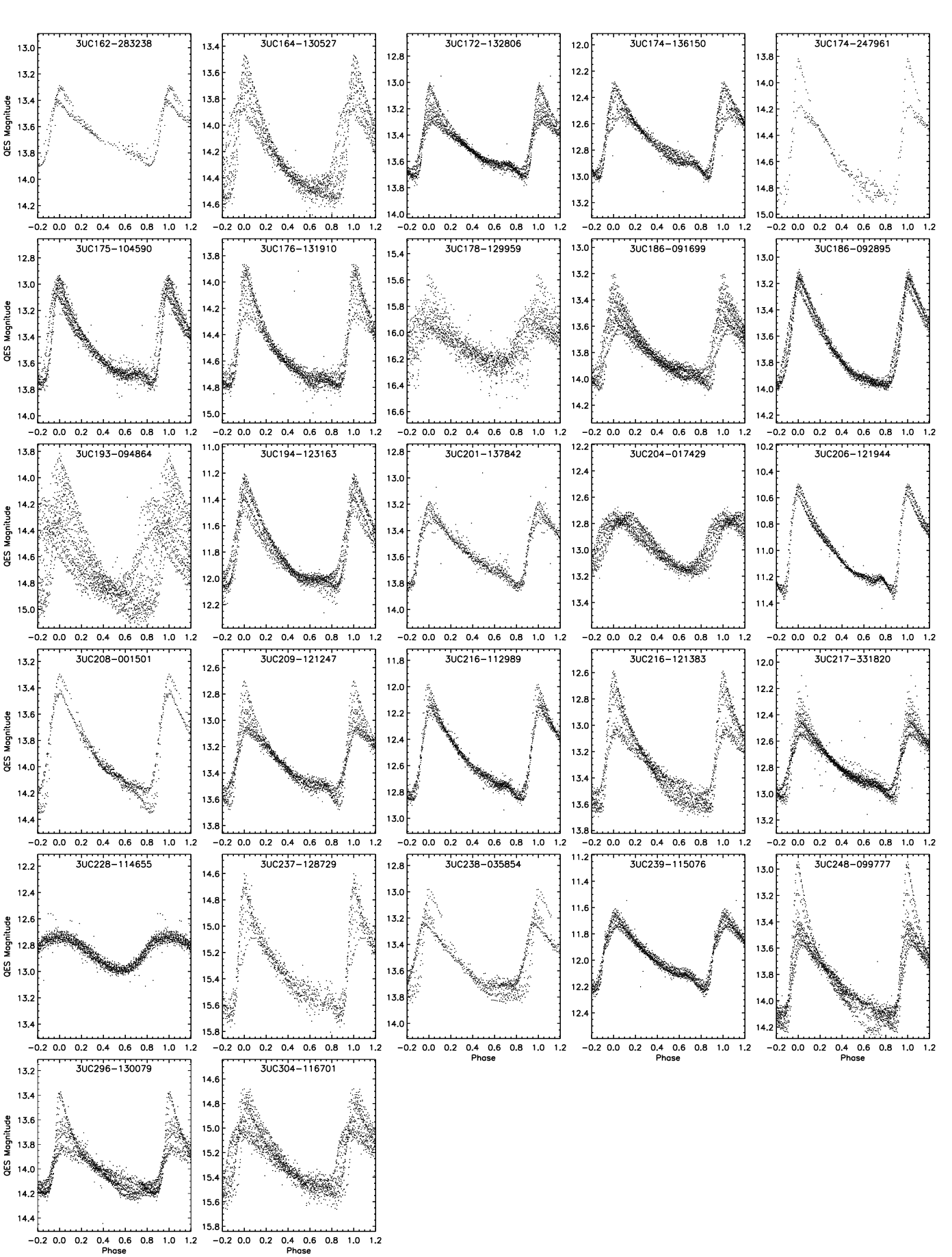}{Phased light curves of the RR Lyrae stars exhibiting the Blazhko effect and that are not in the catalogue of Skarka (2013).
                        The magnitude range in each plot is 1.4~mag.}
\IBVSfigKey{fig8.ps}{Blazhko}{Phased light curves of the RR Lyrae stars exhibiting the Blazhko effect.}

\begin{table}[!ht]
\centerline{{\bf Table\,6.} Time-series photometry for the 588 variable stars described in this paper.}
\centerline{The epoch of mid-exposure (heliocentric Julian date) is listed in column~4. The}
\centerline{magnitudes in column~5 are calibrated QES magnitudes. Column~6 contains the}
\centerline{uncertainties on the magnitudes. The light curve identifier string is listed in}
\centerline{column~7 and consists of a concatenation of the target field name, a camera}
\centerline{identifier and an observing campaign identifier.}
\scriptsize
\begin{center}
\begin{tabular}{@{}ccccccc@{}}
\hline
UCAC3 ID      & GCVS ID   & Variable  & HJD           & $m$    & $\sigma_{m}$ & Light Curve \\
              &           & Type      & (d)           & (mag)  & (mag)        & Identifier  \\
\hline

3UC161-102683 & DH Hya    & RRAB      & 2455639.60565 & 12.265 & 0.256        & 084014$-$044854\_416\_C5 \\
3UC161-102683 & DH Hya    & RRAB      & 2455639.61285 & 12.347 & 0.262        & 084014$-$044854\_416\_C5 \\
\vdots        & \vdots    & \vdots    & \vdots        & \vdots & \vdots       & \vdots                   \\
3UC161-102683 & DH Hya    & RRAB      & 2455639.61110 & 12.268 & 0.377        & 092014$-$044854\_416\_C5 \\
3UC161-102683 & DH Hya    & RRAB      & 2455639.61810 & 12.296 & 0.380        & 092014$-$044854\_416\_C5 \\
\vdots        & \vdots    & \vdots    & \vdots        & \vdots & \vdots       & \vdots                   \\
3UC162-107613 & SZ Hya    & RRAB      & 2455229.77350 & 11.529 & 0.076        & 091000$-$071400\_403\_C2 \\
3UC162-107613 & SZ Hya    & RRAB      & 2455229.78082 & 11.427 & 0.045        & 091000$-$071400\_403\_C2 \\
\vdots        & \vdots    & \vdots    & \vdots        & \vdots & \vdots       & \vdots                   \\
\hline
\end{tabular}
\end{center}
\end{table}

\end{document}